\documentclass[11pt,twoside]{article}
\usepackage{asp2014}

\begin{document}

\title{Overdensities of SMGs around WISE-selected, ultra-luminous, high-redshift galaxies}
\author{Suzy F. Jones,$^1$  and Andrew W. Blain$^1$}
\affil{$^1$Physics $\&$ Astronomy,University of Leicester, Leicester, LE1 7RH, UK; \email{sfj8@le.ac.uk}}

\begin{abstract}
Submillimetre (submm) observations of WISE-selected, dusty, luminous, high-redshift galaxies have revealed intriguing overdensities around them on arcmin scales. They could be the best signposts of overdense environments on the sky.
\end{abstract}
\noindent NASA's Wide-Field Infrared Survey Explorer (WISE) (Wright et al., 2010) found dusty, luminous, high-redshift, active galaxies because the hot dust heated by AGN and/or starburst activity can be traced using the WISE 12\,$\mu$m (W3) and 22\,$\mu$m (W4) bands. Eisenhardt et al. (2012), Bridge et al. (2013) and Lonsdale et al. (submitted) have shown that WISE can find different classes of interesting, luminous, high-redshift, dust-obscured AGN: with faint or undetectable flux densities in the 3.4\,$\mu$m (W1) and 4.6\,$\mu$m (W2) bands, and well detected fluxes in the W3 and/or W4 bands. A radio blind sample known as ``W1W2-dropouts'' or hot dust-obscured galaxies (Hot DOGs) was observed in the submm/mm by Wu et al. (2012) and Jones et al. (2014).  A radio-selected sample known as WISE/radio-selected AGNs have also been observed with JCMT SCUBA-2 at 850\,$\mu$m (Jones et al. 2015) and ALMA (Lonsdale et al. submitted). Submm observations are used to find their coldest dust emission and to be able to calculate their total IR luminosities. Both samples were found to have high total IR luminosities, spectral energy distributions (SEDs) that were not well represented by standard AGN templates, due to excess mid-IR emission to submm ratio (Jones et al. 2014, 2015).

Serendipitous SMG sources are detected in the deepest 850\,$\mu$m SCUBA-2 1.5-arcmin-radius regions around both the Hot DOGs and WISE/radio-selected AGNs. There were 17 and 81 serendipitous 850\,$\mu$m sources detected at greater than 3$\sigma$ significance in the 10 Hot DOG and 30 WISE/radio-selected AGN SCUBA-2 maps, respectively. Both samples were overdensity of accompanying SMGs compared to those in two different blank-field submm surveys (Weiss et al. 2009; Casey et al. 2013).

Observations of dusty, luminous, high-redshift galaxies have revealed significant evidence that the galaxy density in their environments appears to be above average (Blain et al. 2004; Scott et al. 2006; Farrah et al. 2006; Chapman et al. 2009; Hickox et al. 2009; Cooray et al. 2010; Hickox et al. 2012). For example, the Clusters Around Radio-Loud AGN (CARLA) \textit{Spitzer} programme that looked at the environments of radio-loud AGN (RLAGN) at $1.2 < z < 3.2$ concluded that RLAGN are in overdense environments in mid-IR wavelengths, and could be signposts of high-redshift galaxy clusters (Wylezalek et al. 2013; Hatch et al. 2014). Clustering of these mid-IR and SMGs could yield evidence for the nature of their massive dark matter halos and highlight their bias of as compared with the underlying dark matter distribution.

An overdensity of SMGs was found in the fields of Hot DOGs and WISE/radio-selected AGN by factors of $\sim$ 3 and $\sim$ 5, respectively (Jones et al. 2014, 2015). Hot DOGs have a higher typical redshift ($z=2.7$), compared to WISE/radio-selected AGNs ($z=1.7$). Due to the K-correction effect, the serendipitous SMG detection is independent of redshift, and so while the Hot DOGs have a typically higher redshift than the WISE/radio-selected AGNs, the higher overdensity of SMG sources around WISE/radio-selected AGNs is matched in luminosity of the Hot DOG companions. The higher overdensity around WISE/radio-selected AGN could be due to the lower redshift.

This agrees with ALMA cycle-0 results of WISE/radio-selected AGNs (Silva et al. 2014), where 23 serendipitous SMG sources were detected in 17 out of 49 fields, which implies an overdensity factor of $\sim$ 10 on $\sim$ 20\,arcsec scales. There is double the overdensity of SMG serendipitous sources in the ALMA fields than our SCUBA-2 fields when compared to previous submm surveys, this could be due to the multiplicity in SMGs (Karim et al. 2013). This is where multiple SMGs are found to be separated typically by $\sim$ 6\,arcsec and are resolved with high-resolution ALMA maps (1.5\,arcsec resolution), but are blended into one source with the resolution of single-dish surveys, where JCMT used in this paper has a resolution of 15\,arcsec at 850\,$\mu$m. There were no fields in both samples that contained 0 serendipitous sources, with $\sim$ 2 typically found in each Hot DOG field and $\sim$ 3 in WISE/radio-selected AGNs compared to $\sim$ 1 typically found in blank field submm surveys (Coppin et al. 2006; Weiss et al. 2009; Casey et al. 2013).

Hot DOGs and WISE/radio-selected AGNs are highly luminous and could be the best signposts of overdense environments of active, dusty, luminous galaxies.


\vspace{0.2cm}
\noindent References:

\noindent Blain A. W. et al., 2004, ApJ, 611, 725   

\noindent Bridge C. R. et al., 2013, ApJ, 769, 91 

\noindent Casey C. M. et al., 2013, MNRAS, 436, 1919

\noindent Chapman S. C. et al., 2009, ApJ, 691, 560 

\noindent Coppin K. et al., 2006, MNRAS, 372, 1621

\noindent Cooray A. et al., 2010, A$\&$A, 518, L22

\noindent Eisenhardt P. R.. et al., 2012, ApJ, 755, 173

\noindent Farrah D. et al., 2006, ApJL, 641, L17 

\noindent Hatch N. A. et al., 2014, MNRAS, 445, 280

\noindent Hickox R. C. et al., 2009, ApJ, 696, 891

\noindent Hickox R. C. et al., 2012, MNRAS, 421, 284

\noindent Jones S. F. et al., 2014, MNRAS, 443, 146

\noindent Jones S. F. et al., 2015, MNRAS, ArXiv 1503.02561

\noindent Karim A. et al., 2013, MNRAS, 432, 2

\noindent Scott S. E. et al., 2006, MNRAS, 370, 1057

\noindent Silva A. L. et al., 2014, in American Astronomical Society Meeting Abstracts, 224

\noindent Tsai C. W. et al., 2014, ArXiv 1410.1751T

\noindent Weiss A. et al., 2009, ApJ, 707, 1201 

\noindent Wright E. L. et al., 2010, AJ, 140, 1868 

\noindent Wu J. et al., 2012, ApJ, 756, 96

\noindent Wylezalek D. et al., 2013, ApJ, 769, 79

\end{document}